\documentstyle[11pt,aasms]{article}
\input epsf
\begin{document}
\title{Determining cosmic microwave background structure from its peak
distribution.}

\author{A. Kashlinsky}
\affil{Raytheon ITSS\\ Code 685, Goddard Space Flight Center, Greenbelt,
MD 20771\\
	e--mail: kashlinsky@stars.gsfc.nasa.gov}

\author{C. Hern\'andez--Monteagudo}
\affil{F{\'\i}sica Te\'orica. Facultad de Ciencias.\\
	Universidad de Salamanca, 37008 Spain.\\
	e--mail: chm@orion.usal.es}

\author{F. Atrio--Barandela}
\affil{F{\'\i}sica Te\'orica. Facultad de Ciencias.\\
	Universidad de Salamanca, 37008 Spain.\\
	e--mail: atrio@orion.usal.es}

\begin{abstract}
We present a new method for time-efficient and accurate extraction of the power
spectrum from future cosmic microwave background (CMB) maps based on properties
of peaks and troughs of the Gaussian CMB sky. We construct a statistic
describing their angular clustering - analogously to galaxies, the 2-point
angular correlation function, $\xi_\nu(\theta)$. We show that for increasing
peak threshold, $\nu$, the $\xi_\nu(\theta)$ is strongly amplified and becomes
measurable for $\nu\geq$1 on angular scales $\leq 10^\circ$. Its amplitude at
every scale depends uniquely on the CMB temperature correlation function,
$C(\theta)$, and thus the measured $\xi_\nu$ can be uniquely inverted to
obtain $C(\theta)$ and its Legendre transform, the power spectrum of the CMB
field. Because in this method the CMB power spectrum is deduced from high
peaks/troughs of the CMB field,
the procedure takes only $[f(\nu)]^2N^2$ operations where $f(\nu)$ is the
fraction of pixels with $|\delta T|\geq\nu$ standard deviations in the map of
$N$ pixels and is e.g. 0.045 and 0.01 for $\nu$=2 and 2.5 respectively.
We develop theoretical formalism for the method and show with detailed
simulations, using MAP mission parameters, that this method allows to determine 
very accurately the CMB power
spectrum from the upcoming CMB maps in only $\sim(10^{-4}-10^{-3})\times N^2$
operations.

\end{abstract}
{\bf Subject headings}: cosmology - cosmic microwave background - methods: 
numerical
\newpage
{1. \bf Introduction.}

By probing the structure of the last scattering surface, the current and
upcoming balloon and space borne missions promise to
revolutionize our understanding of the early Universe physics. This
requires probing the angular spectrum of the cosmic microwave background (CMB)
with high precision on sub-degree scales, or angular wave-numbers $l>
200$. For cold-dark-matter (CDM) models based on
inflationary
model for the early Universe and adiabatic density perturbations, the
structure of the CMB should show the signature of acoustic oscillations leading
to multiple (Doppler) peaks. The relative spacing of the Doppler peaks should
then reflect the overall geometry of the Universe, whereas the amplitude of the
second (and higher) peaks depends sensitively on other cosmological parameters,
such as the baryon density, re-ionization epoch, etc. The recent balloon-borne
measurements (de Bernardis et al.,
2000, Mauskopf et al 2000, Hanany et al., 2000) strongly imply a flat
cosmological model because the first Doppler peak occurs at $l \simeq 200$ 
(Kamionkowski, Spergel \&
Sugiyama 1994, Melchiori et al 2000, Jaffe et al
2001).

A major challenge to understanding these and future measurements is to find an
efficient
algorithm that can reduce the enormous datasets with $N\!\simeq\!10^5$ pixels 
in
balloon
experiments to $\simeq\!3\times\! 10^6$ for MAP band with the 0.2$^\circ$ beam 
(at
90 GHz)
to $\simeq\!10^8$ for the Planck HFI data. Traditional methods require
inverting the covariance matrix and need $\sim\!N^3$ operations making them
impossible for the current generation of computers. 
Thus alternatives have been
developed for estimating the CMB multipoles in O($N^2$) operations from 
Gaussian sky
(Tegmark 1997; Oh, Spergel \& Hinshaw 1999) and general CMB sky (Szapudi et al 
2000) 
as well as study statistics of the various methods (Bond et al. 2000;
Wandelt et al 2001).

In this {\rm letter} we suggest a novel, accurate and time-efficient method for
computing the angular power spectrum of the CMB temperature field from datasets
with large $N$ using the properties of the peaks (and troughs) of the CMB
field.
The peaks are much fewer in number than $N$, but they will be strongly
clustered.
In Sec. 2 we construct a statistic describing their angular clustering - the
2-point
angular correlation function $\xi$ in analogy to the galaxy correlation
function
(Peebles 1980), i.e. the excess probability of finding two peaks at a given
separation
angle. We show that this statistic is strongly amplified over the
scales of interest ($<$10$^\circ$) and should be measurable. The value of
$\xi$
for a
given peak threshold $|\delta T| \!\geq \!\nu \sigma$ would be uniquely related 
to
the
correlation function of the temperature field $C\! = \!\langle \delta T(\vec x)
\delta
T(\vec x \!+ \!\vec \theta)\rangle$. The measurement of $\xi$ can then be 
uniquely
inverted to obtain the underlying $C$ and its Fourier transform, the power
spectrum
$C_l$. This can be achieved in just $[f(\nu)]^2 N^2$ operations, where $f(\nu)$
is the
fraction of pixels with $|\delta T| \!\geq \!\nu \sigma$ and is e.g. 4.5--1\% 
for
$\nu$=2-2.5. Sec. 3 shows concrete numerical simulations for the MAP
90GHz channel in order to estimate cosmic variance,
sampling uncertainties, instrumental noise etc. We show that with this method
the
CMB power spectrum is recovered
to accuracy comparable to or better than by other existing methods but in a
significantly smaller number of operations. On an UltraSparc II 450 MHz
processor the entire sky
map with 0.2$^\circ$ angular resolution can be analyzed and $C_l$'s recovered in
only 15 mins and 2.25 hours CPU time for $\nu=$2.5 and 2.1 respectively. 

{2. \bf Method}

For Gaussian ensemble of $N$ data points (e.g. pixels) describing the CMB data
$\delta \equiv T - \langle T \rangle$ one expects to find a fraction
$f(\nu)=$erfc($\nu/\sqrt{2})$ with $|\delta|
\geq
\nu\sigma$, where $\sigma^2=\langle \delta^2\rangle$ is the variance of the
field
and erfc is the complementary error function. E.g.
$f(\nu)$=$(4.5,1,0.1)\!\times\!
10^{-2}$ for $\nu$=(2,2.5,3) respectively.
The joint probability density of finding two pixels
within $d\delta_{1,2}$ of $\delta_{1,2}$ and separated by the angular distance
$\theta$ is given by the bivariate Gaussian (the vector 
$\mbox{\boldmath$\delta$}$ has components [$\delta_1,\delta_2$]):
\begin{equation}
p(\delta_1,\delta_2) = \frac{1}{2\pi \sqrt{||\mbox{\boldmath$C$}||} }
\exp(-\frac{1}{2} \mbox{\boldmath$\delta$} \cdot  \mbox{\boldmath$C$}^{-1}
\cdot  \mbox{\boldmath$\delta$}) =
\frac{1}{(2\pi)^2} \int_{-\infty}^{\infty}
\int_{-\infty}^{\infty}
\exp(-i \mbox{\boldmath$q$} \cdot  \mbox{\boldmath$\delta$})
\exp(-\frac{1}{2} \mbox{\boldmath$q$} \cdot  \mbox{\boldmath$C$}
\cdot  \mbox{\boldmath$q$}) d^2\mbox{\boldmath$q$}
\label{P_definition}
\end{equation}
where $C$ is the covariance matrix of the temperature field. We model the
covariance
matrix in (\ref{P_definition}) as
$C(\theta) = C_0\delta_{ij} + C(\theta_{ij})(1 - \delta_{ij})$,
where $\delta_{ij}$ is the Kronecker delta and
$C_0 \equiv C(0) + \sigma_n^2$; $\sigma_n$ is the noise contribution. We assume
that
the noise is diagonal, the entire
CMB sky is Gaussian and the cosmological signal whose power spectrum we seek is
contained in $C(\theta) = \sum (2l+1)C_l P_l(\cos\theta)/4\pi$. The total
dispersion
of the temperature field is then $\sigma=\sqrt{C(0)+\sigma_n^2}$

The peaks of a Gaussian field should be
strongly clustered (Rice 1954, Kaiser 1984, Jensen \& Szalay 1986, Kashlinsky
1987). The
angular clustering of such regions can be described by the 2-point
correlation
function, i.e. the excess probability of finding two events at the given
separation.
The probability of simultaneously finding two temperature excursions with
$|\delta T|\! \geq \!\nu \sigma$ in small solid angles $dw_{1,2}$ is $dP_{12}
\!\propto\!
(1+\xi)dw_1 dw_2$.  The correlation function of such regions is:
\begin{equation}
\xi_\nu(\theta) = \frac{ 2 \int_{\nu\sigma}^\infty\int_{\nu\sigma}^\infty
[p(\delta_1,\delta_2)+p(-\delta_1,\delta_2)] d\delta_1d\delta_2}
{[2\int_{\nu\sigma}^\infty p(\delta) d\delta]^2} - 1
\label{xi_definition}
\end{equation}
The numerator follows from considering contributions from correlations between
$\delta_1, \delta_2$ in regions of 1) $\delta_1 \!\geq\! \nu\sqrt{C_0}$,
$\delta_2\! \geq\! \nu\sqrt{C_0}$; 2) $\delta_1\! \leq\! - \nu\sqrt{C_0}$, 
$\delta_2\!
\leq \!- \nu\sqrt{C_0}$;
and 3) twice the contribution of $\delta_1\! \geq\! \nu\sqrt{C_0}$,
$\delta_2\! \leq\! - \nu\sqrt{C_0}$. The ``2" in the denominator comes 
because we consider both peaks and troughs.

In order to evaluate (\ref{xi_definition}) directly, we expand
$\exp[-q_1q_2C(\theta)]
=\sum_{k=0}^\infty \frac{[-C(\theta)]^k}{k!} q_1^k
q_2^k$ in (1) and use the fact that
$\int_{-\infty}^\infty \exp(-ixy) F(x) x^k dx=
i^k(\partial^k/\partial y^k) \int_{-\infty}^\infty \exp(-ixy) F(x) dx$
(Jensen \& Szalay 1987, Kashlinsky 1991).
Because
$\int_{-\infty}^\infty \exp(-iq\delta) \exp(-q^2C_0/2) dq=
\sqrt{2\pi/C_0} \exp(-\frac{\delta^2}{2C_0})$
we get:
\begin{equation}
p(\delta_1,\delta_2) = \frac{1}{2\pi C_0} \sum_{k=0}^\infty
\frac{[C(\theta)]^k}{k!}
\left[\frac{\partial^k }{\partial \delta_1^k}\exp(-\frac{\delta_1^2}{2C_0})
\right]
\label{P_hermite}
\left[\frac{\partial^k }{\partial \delta_2^k}\exp(-\frac{\delta_2^2}{2C_0})
\right]
\label{p12}
\end{equation}
Substituting (\ref{P_hermite}) into (\ref{xi_definition}) allows to expand
$\xi_\nu(\theta)$ into the Hermite
polynomials, $H_n(x)=(-)^n \exp(x^2) (d^n/dx^n)\exp(-x^2)$, to obtain:
\begin{equation}
\xi_\nu(\theta) = A_\nu(\frac{C}{C_0})
\label{xi}
\end{equation}
with:
\begin{equation}
A_\nu (x)=\frac{1}{H^2_{-1}(\frac{\nu}{\sqrt{2}})}
\sum_{k=1}^\infty \frac{ x^{2k} }{ 2^{2k}(2k)!
}H_{2k-1}^2(\frac{\nu}{\sqrt{2}})
\label{amplification}
\end{equation}
where $H_{-1}(x)\!\equiv \!\frac{ \sqrt{\pi} }{2}
\exp(x^2)$erfc($x)$.
At each angular scale the value of $\xi_\nu$ for every $\nu$ is determined
uniquely by $C$ at the same $\theta$. Note that in the limit of the entire
map ($\nu$=0) our statistic $\xi_\nu$=0 and our method becomes meaningless; the
new statistic has meaning only for sufficiently high $\nu$. One should 
distinguish between the 2-point correlation function, $\xi$, we directly 
determine from the maps, and the commonly used statistics in CMB studies, the 
temperature correlation function, $C$.

Fig. 1 shows the properties of $\xi_\nu$: the left panel shows the variation of
$\xi_\nu$ with $C/C_0$ for fixed $\nu$ and the middle panel shows
the variation of $\xi_\nu$ with $\nu$ for fixed $C/C_0$.
The first term in the sum in eq. (\ref{amplification}) contains $H_1^2 (\propto
\!\nu^2$), but as the middle panel of Fig.1 shows
for sub-degree scales (where $|C/C_0|\!>\!0.1$) $\xi_\nu$ changes more steeply 
than 
$\nu^2$. This means that $k\!>$1 terms are important for accurate inversion of
$\xi_\nu$ in terms of $C_l$'s. The right panel shows $\xi_\nu$ vs the angular 
separation $\theta$ for
$\nu$=2 for two flat CDM models: $\Lambda$CDM model (thin line) with 
$(\Omega_{\rm total}, \Omega_\Lambda)$=(1,0.7) and $\Omega_{\rm
baryon}h^2$=0.03 and SCDM with $(\Omega_{\rm total}, \Omega_\Lambda)$=(1,0) and
$\Omega_{\rm baryon}h^2$=0.01 (thick line). The first model
has prominent Doppler peaks and baryon
abundance in agreement with BBNS, the second model requires significantly
higher
baryon abundance but has a much smaller second Doppler peak. 
There 
would be non-linear to
quasi-linear
(and easily detectable) clustering of
high peaks out to the angular scale where $C(\theta)$ drops to only $\sim$0.1
of
its
maximal value at zero-lag. This covers the angular scales of interest for
determining the sub-horizon structure at the last scattering. Because the
uncertainty in measuring $\xi$ is $\sim \!N_{\rm pairs}^{-1/2}$ (Peebles 1980),
the
value of $\xi$ can
be determined quite accurately in non-linear to quasi-linear regime. At the
same time, as the left panel in Fig.1 shows, over this range of scales the
amplitude of
$\xi_\nu$ changes rapidly with $C$ making possible a stable inversion
procedure
to obtain $C(\theta)$ from $\xi_\nu$.

This suggests the following procedure to determine the power spectrum
of CMB in only $\simeq\!f^2(\nu)N^2$ operations:
$\bullet$ Determine the variance of the CMB
temperature, $C_0$, from the data in $N$ operations;
$\bullet$ Choose sufficiently high $\nu$ when $f(\nu)$ is small
but at the same time
enough pixels are left in the map for robust measurement of
$\xi_\nu(\theta)$;
$\bullet$ Determine $\xi_\nu(\theta)$ in $[f(\nu)]^2 N^2$
operations.
$\bullet$ Finally, given the values of $(C_0,\nu)$ solve equation
$A_\nu(C/C_0)$=$\xi_\nu(\theta)$ to obtain $C(\theta)$ and from it $C_l$.

{3. \bf Numerical results and applications}

In order to apply the proposed method in practice and to estimate the cosmic
variance,
sampling and other uncertainties, we ran numerical simulations with
parameters
corresponding to the MAP 90 GHz channel. The CMB sky was simulated using
HEALPix
(G\'orski et al 1998) software with $N_{\rm side}$=512 and Gaussian beam with
FWHM=0.21$^{\circ}$ (or $l_{\rm Nyquist}$=640) for the SCDM and $\Lambda$CDM 
models. To this we added Gaussian
white noise with the rms of 35$\mu$K per
0.3$^\circ\!\times\! 0.3^\circ$ pixel (Hinshaw 2000). We assumed that the
foreground contribution at 90 GHz can be subtracted
to within a negligible term. Fig. 2 shows the SCDM
model sky with both peaks and troughs with $|\delta|\! \geq\! \nu\sqrt{C_0}$, 
$\nu$=2,
marked
with white dots. The clustering of peaks/troughs is very prominent especially
on
small scales, but the clustering pattern is very different
between the models.

To determine $\xi_\nu$ we divided $\theta$ into 31415 equally spaced bins 
between
0 and 180$^\circ$ and the number of pixel pairs, $N_{12}$, with 
$|\delta|\!\geq\! \nu\sigma$, was oversampled and determined 
in each bin. This is the dominant
CPU time-consuming procedure of $[f(\nu) N]^2$ operations. 
The ''raw" value of $\xi_{\rm raw}$ was 
determined in each of the 31415 bins as $\xi_{\rm raw}$=$ N_{12}/N_{rr}-1$, 
where $N_{rr}$ is the number of pairs for a Poissonian catalog with the total 
number of pixels equal to the number of peaks and troughs in our 
CMB sky. The final
$\xi$ was determined as follows: the angular interval between 0 and 180$^{\rm
o}$ was divided into bins centered on the roots of the 800th order Legendre
polynomial in order to facilitate the later inversion of
$C(\theta)$ into $C_l$'s via the Gauss-Legendre integration. The final $\xi$
was obtained from $\xi_{\rm raw}$ by
convolving the latter with a Gaussian filter of 4' dispersion centered
on each of the 800 Legendre polynomial roots. The value of thus obtained
$\xi(\theta)$ is shown for one realization of the two CDM models in the right
panel of Fig.1; it agrees well with eqs. 
(\ref{xi},\ref{amplification}).

Having fixed $\nu$ and determined $C_0$ from the map we now solve the equation
$\xi_\nu(\theta)$=$A_\nu(C(\theta)/C_0)$ with respect to $C(\theta)$ with 
$A_\nu$
given by eq.(\ref{amplification}) and determine $C(\theta)$ at each
of the roots
of $l$=800 Legendre polynomial. In the final step the multipoles were 
determined by direct Gauss-Legendre
integration of $C_l$=$ 2\pi \int C(\theta ) P_l (\cos\theta )\sin \theta
d\theta$. At $\theta\!>$10$^\circ$, where $\xi_\nu$ is very small and hard to 
determine (see Fig.1c), our recovered $C(\theta)$ has larger uncertainties. 
Because we are interested in high $l$ multipoles the recovered 
correlation function $C(\theta )$ was further tapered above 15$^\circ$. (At 
$l$ of interest the results are insensitive to details of tapering). 
To check the statistical uncertainties in the
determination of
$C_l$'s we ran 700 simulations for $\nu$=2.5 and 350 simulations for $\nu$=2.1.

Fig. 3 shows the results of the numerical simulations for SCDM model with the
instrument noise of the MAP 90 GHz channel. The distribution of the multipoles
determined from the simulated maps in this method is shown in Fig. 3a,b for
$l$=200 (the first Doppler peak), $l$=350 (the first trough) and $l$=475 (the 
second Doppler 
peak for SCDM model). The best-fit Gaussians are shown with smooth lines; they 
give good fit to the histograms at high $l$. The 68\% confidence limits on 
$C_l$'s are very close to the dispersion of the best-fit Gaussians shown 
and the 95\% limits are roughly twice as wide for all $l$'s as would
be the case for approximately Gaussian distributions.

Fig. 3c shows $C(\theta)$ determined by our method from one realization for
$\nu$=2.1 (solid line) and $\nu$=2.5 (dashes). Dotted line shows the
theoretical
$C(\theta)$=$\sum (2l+1)C_l P_l(\cos\theta)/4\pi$ with SCDM values of $C_l$'s.
The $C(\theta)$ determined with our method is within 5\% of
the theoretical value. This uncertainty is within the cosmic variance of
small-scale $C(\theta)$ from low-$l$ contribution (predominantly quadrupole)
which
is $\sim 500 \mu$K$^2$ (Bennett et al 1996, Hinshaw et al 1996).

Finally, Fig. 3d shows our full sky determination of the CMB power spectrum from 
the synthesized SCDM maps. Solid line shows the theoretical (input) spectrum 
juxtaposed with the power spectrum determined with the peaks
method for $\nu$=2.1 (diamonds) and 2.5 (crosses). The latter was band-averaged 
into
$\Delta l$=50 wide bins and the symbols are plotted at the central bin value
+/-- 5 for $\nu$=2.5/2.1 respectively to enable clearer
display. Band power averaged over larger $\Delta l$ will have
less variance but will give fewer independent data points. In computing the 
band-averages we gave equal weight to all
multipoles. This top-hat window function is not an optimal estimator
of the power spectrum (Knox 1999). However, we checked that 
with this window function, the multipoles at the different 
$l$-bins were not correlated. 
The shown uncertainties correspond to the dispersion from the 
Gaussian fits to the distributions and are very close to the 68\% error bars 
which are approximately half 
the 95\% errors. With our method for $\nu$=2.1 we recover the
power spectrum with variance only 1\% larger than the full-sky cosmic variance 
limit at the central $l$ at 
the first Doppler peak ($l$=200), 2\% larger at $l$=350 and 8\% larger at the 
second peak ($l$=475). We compare the uncertainty of our numbers to the full-sky 
cosmic variance which is, excluding the noise, $\Delta C_l$=$\sqrt{2/(2l+1)} 
C_l$. Note 
that for our method 
we simulated the two-year full-sky MAP 
90 GHz channel, so our assumed noise is $\sim$100$\mu$K per 7' 
pixel with the FWHM=12.6' beam, i.e. twice that assumed in 
Szapudi et al (2000) with the FWHM=10' beam who obtain a similar accuracy 
with direct computation of $C(\theta)$ for a small BOOMERANG-size patch of the 
sky. At $l\!\leq\!l_{\rm Nyquist}$ of the 
beam our method determines $C_l$'s without bias. The accuracy of our method for 
a given $\Delta l$ can be further improved by going to $\nu\!<$2.1, but 
increasing somewhat computational time (e.g. for $\nu$=2 the computational time 
will increase by 60\% over $\nu$=2.1).

{4. \bf Conclusions}

We introduced here a new statistic, $\xi_\nu$, and showed that with it one can 
recover the power 
spectrum from Gaussian CMB maps in a very accurate and time-efficient way. We 
have shown that for peaks and troughs of 
such temperature field, their angular 2-point correlation function, $\xi_\nu$, 
would be 
strongly amplified with increasing threshold $\nu$ and would be measurable. 
Because its amplitude at a given angular scale depends on the amplitude of the 
temperature correlation function, $C$, at the same scale, the former is then 
inverted to obtain the power spectrum of the CMB. The method requires 
$[f(\nu)N]^2\!\ll\! N^2$ operations.  For balloon experiments it would work for 
$\nu\!\sim$1--1.5 or in $(10^{-1}\!-\!10^{-2})N^2$ operations, for MAP highest 
resolution with $\sim \!3\!\times \!10^{6}$ pixels it can work at 
$\nu\!\simeq$2--2.5 or 
in $(2\!\times\! 10^{-3}\!-\!1.5 \times 10^{-4})N^2$ operations and for higher 
resolution 
maps, such as Planck HFI maps, still higher $\nu$ can be used leading to 
accurate results in $<\! 10^{-4}N^2$ operations. We demonstrated with 
simulations 
that for the two-year noise levels for MAP 90GHz channel with this method we 
can 
recover the CMB multipoles with $\nu$=2.1, or in $1.2\!\times \!10^{-3}N^2$ 
operations, out to $l_{\rm Nyquist}$ with uncertainty only a few percent larger 
than the full-sky cosmic variance. We assumed a diagonal noise covariance 
matrix, 
but the method can be extended to another Gaussian noise 
provided its covariance matrix is known. Because here we work with the 
correlation functions, our method is immune to geometrical masking effects e.g. 
from Galactic cut and other holes in the maps. This means that we can remove 
regions where the foregrounds are very bright, or the noise inhomogeneity is too 
high. With the exception of isolated 
areas, the MAP noise variations are expected to be $\sim$10--15\% over the cut 
sky (Hinshaw, private communication). Because $\sigma_n^2\!\ll \! C(0)$ such 
variations would lead to only small variations in the effective $\nu$. The 
method applies to Gaussian CMB sky 
which is expected in conventional models. If the CMB sky turns out to be 
non-Gaussian, our method may not be applicable, or may need substantial 
modifications by e.g. looking at $\xi_\nu$ for peaks and troughs separately. 

We acknowledge fruitful conversations with Gary Hinshaw and thank Keith 
Feggans at NASA GSFC for generous computer advice and resources. C.H.M. and
F.A.B. acknowledge support of Junta de Castilla y Le\'on (project
SA 19/00B) and Ministerio de Educaci\'on y Cultura (project BFM2000-1322).

{\bf REFERENCES}\\
Bennett, C. et al 1996, Ap.J., 464, 1\\
de Bernardis et al. 2000, Nature, 404, 955\\
Bond, J.R., Jaffe, A.H. \& Knox, L. 2000,Ap.J., 533, 19.\\
G\'orski, K.M., Hivon, E. \& Wandelt, B.D. 1999 in Proc.
MPA/ESO Conf. (eds. Banday, A.J., Sheth, R.K. \& Da Costa, L.)
{\it (website: http://www.eso.org/~kgorski/healpix)}\\
Hanany, S. et al., 2000, Ap.J., 545, L5\\
Hinshaw, G. et al. 1996, Ap.J., 464, L25\\
Hinshaw, G. 2000, astro-ph/0011555\\
Jaffe, A.H. et al. 2001, astro-ph/0007333\\
Jensen, L.G. \& Szalay, A.S. 1986 ApJ, 305, L5\\
Kaiser, N. 1984 Ap.J., 282, L9\\
Kamionkowski, M., Spergel, D.N. \& N. Sugiyama 1995, ApJ 426, L57\\
Kashlinsky, A. 1987, Ap.J. 317, 19\\
Kashlinsky, A. 1992, Ap.J., 386, L37\\
Kashlinsky, A. 1998, Ap.J., 492, 1\\
Knox, L. 1999, Phys. Rev. D, 60, 103516\\
Mauskopf, P.D. et al 2000, Ap.J., 536, L59\\
Melchiori, A. et al 2000, Ap.J., 536, L63\\
Oh, S.P., Spergel, D.N. \& Hinshaw, G. 1999 ApJ, 510, 551\\
Peebles, P.J.E. 1980 {The Large Scale Structure
of the Universe}, Princeton, Princeton University Press\\
Rice, S.O. 1954, in ``Noise and Stochastic Processes", ed. Wax, N., p.133 Dover
(NY)\\
Szapudi, I. et al 2000,Ap.J., 548, L115. (astro-ph/0010256)\\
Tegmark, M. 1997, Phys.Rev.D., Phys.Rev. D, 55, 5898\\
Wandelt, B., Hivon, E. \& G\'orski, K. 2001, Phys.Rev.D, submitted
(astro-ph/0008111)\\
\newpage

FIGURE CAPTIONS

{\bf Fig. 1}: (a) $\xi_\nu$ vs $C/C_0$ for $\nu=0.5,1,1.5,2,2.5,3$ from bottom
to top. (b) $\xi_\nu$ vs $\nu$ for $C/C_0=0.1,0.2,0.3,0.5,0.75,0.95$ from
bottom
to top. (c)  $\xi_\nu$ vs $\theta$ for $\nu=2$ in one realization of the two
CDM
models: Plus signs correspond to $\xi$ determined directly from one simulated
map
of $\Lambda$CDM with FWHM=0.21$^\circ$ resolution and the noise corresponding
to the 90 GHz MAP channel; diamonds show the same for SCDM. Thick and thin
solid
lines show the values of $\xi(\theta)$ from eqs. (\ref{xi},\ref{amplification})
for SCDM and $\Lambda$CDM respectively.

{\bf Fig. 2}: All sky distribution of pixels with $\nu$=2 for SCDM model.

{\bf Fig. 3}: (a), (b) Histograms of the recovered $C_l$'s for $l$=200, 350 and 
475 are
shown for $\nu$=2.5 (top) and 2.1 (bottom). Smooth lines show the best-fit 
Gaussians to the histogram data. (c) $C(\theta)$ vs
$\theta$ for SCDM model: theoretical value is shown with dotted line. The
values for one
realization are shown with solid (for $\nu$=2.1) and dashed ($\nu$=2.5) lines.
(d) $C_l$ vs $l$ for SCDM model. Solid line corresponds to the
theoretical input value. The spectrum recovered with our method from simulated
90
GHz MAP maps is shown after band-averaging with $\Delta l$=50 with filled
diamonds ($\nu$=2.1) and crosses ($\nu$=2.5). To enable a clearer display the
central values of multipoles are shifted by 5 to the left for $\nu$=2.1 and to
the right for $\nu$=2.1. The error bars correspond to the dispersion of the 
Gaussian  fits such as as shown in Fig.3a and practically coincide with 68\% 
confidence limits which in turn are approximately half the 95\% limits.

\newpage
\clearpage
\begin{figure}
\centering
\leavevmode
\epsfxsize=1.0
\columnwidth
\epsfbox{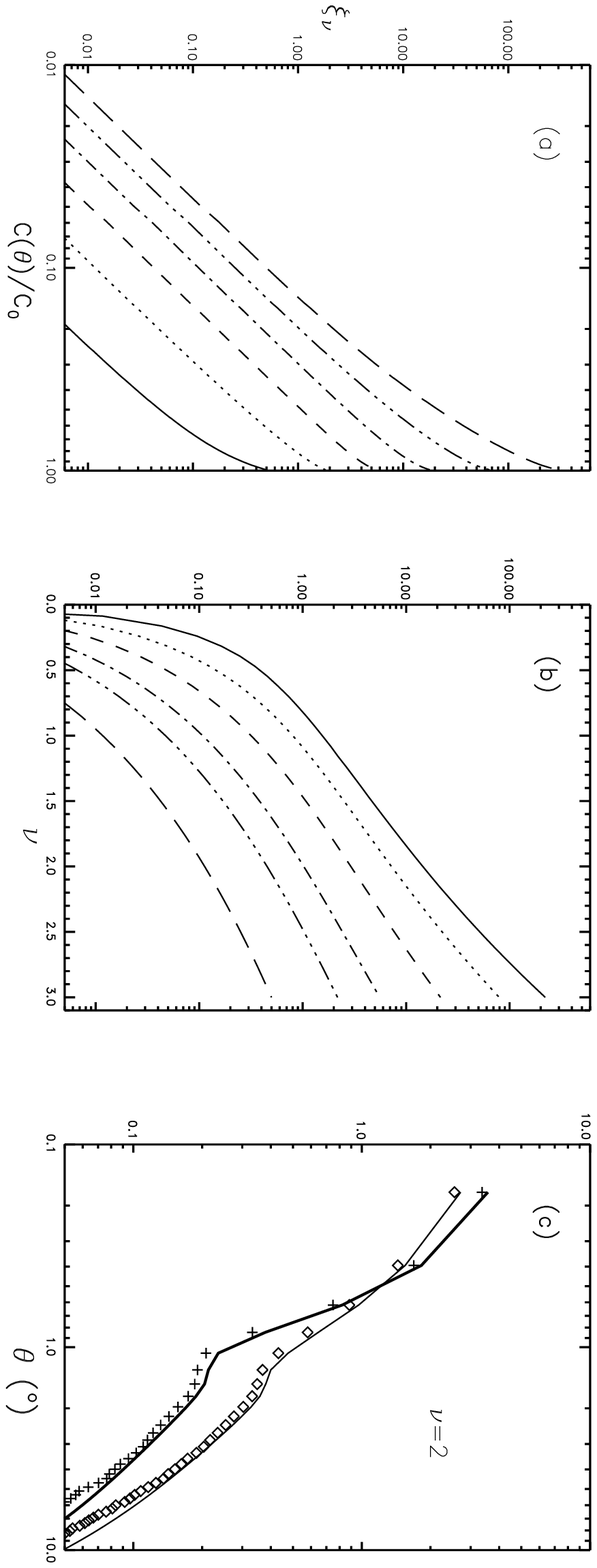}
\caption[]{ }
\label{f1}
\end{figure}

\newpage
\clearpage
\begin{figure}
\centering
\leavevmode
\epsfxsize=0.75
\columnwidth
\epsfbox{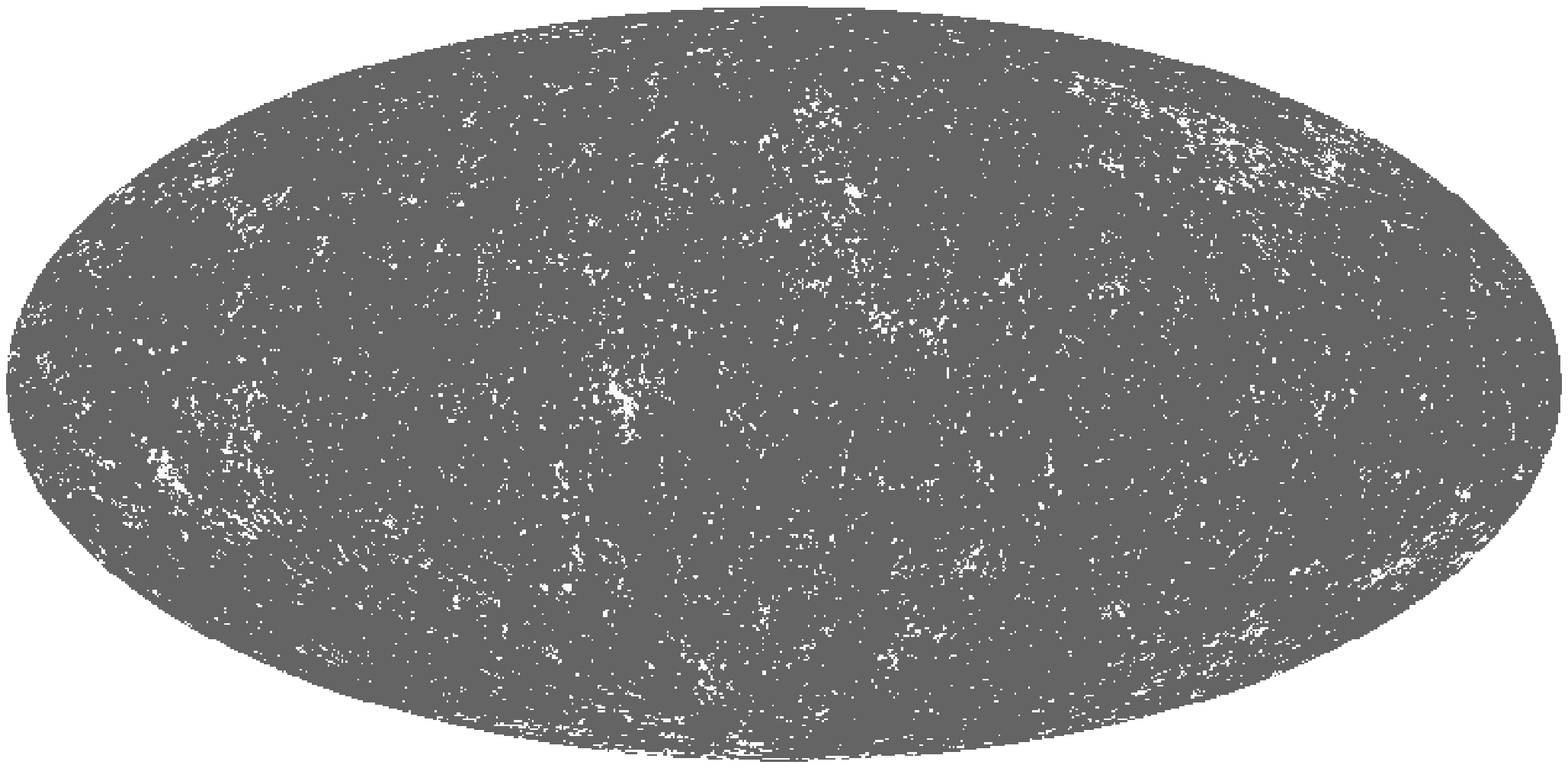}
\caption[]{ }
\label{f2}
\end{figure}

\newpage
\clearpage
\begin{figure}
\centering
\leavevmode
\epsfxsize=0.55
\columnwidth
\epsfbox{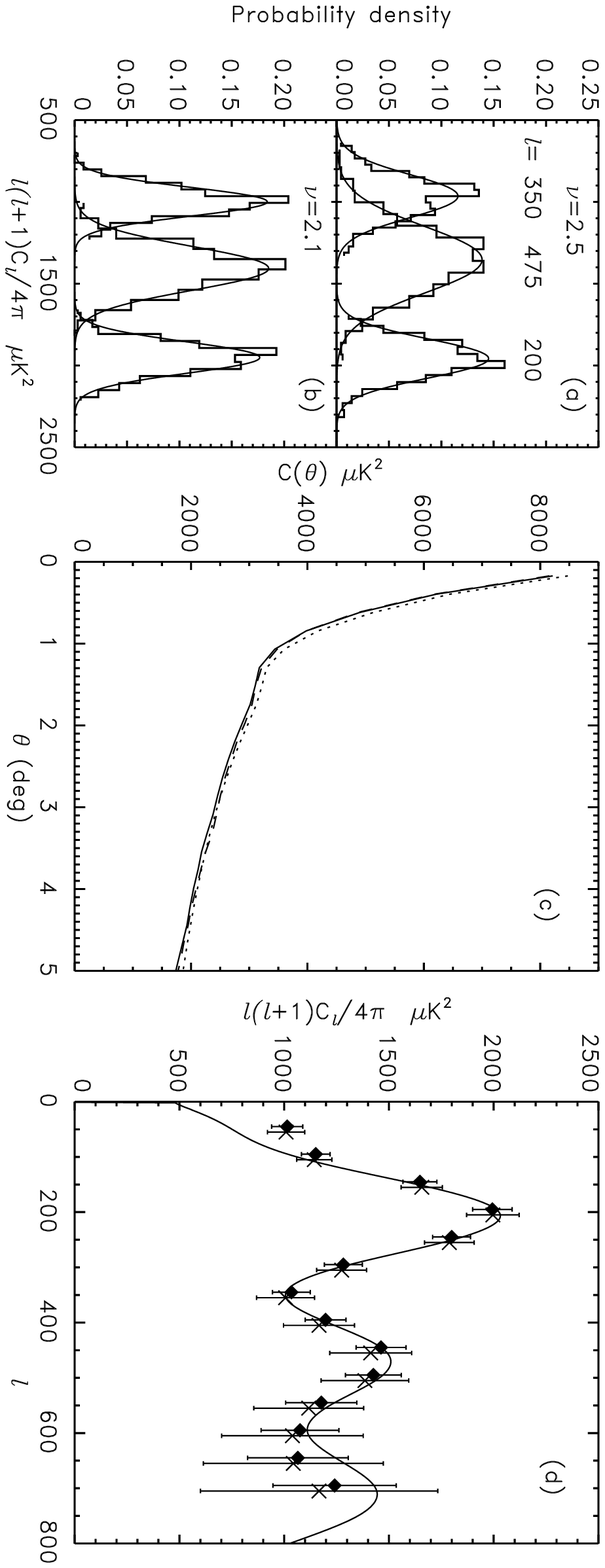}
\caption[]{ }
\label{f3}
\end{figure}


\end{document}